\documentclass[twocolumn,showpacs,preprintnumbers,amsmath,amssymb,aps,prl]{revtex4}
\usepackage{graphicx}
\begin{document}

\title{Dynamics, Rectification,
and Fractionation for Colloids on Flashing Substrates 
} 
\author{A. Lib{\' a}l$^{1,2}$, C. Reichhardt$^{1}$, 
B. Jank{\' o}$^{2}$, and 
C.J. Olson Reichhardt$^{1}$}
\affiliation{ 
{$^1$}Center for Nonlinear Studies and Theoretical 
Division, 
Los Alamos National Laboratory, Los Alamos, New Mexico 87545\\
{$^2$}Department of Physics, University of Notre Dame, Notre Dame, 
Indiana 46556
}
\date{\today}

\begin{abstract}
We show that a rich variety of dynamic phases 
can be realized for 
mono- and bidisperse mixtures
of interacting 
colloids under the influence of a
symmetric flashing periodic substrate.
With the addition of dc or ac drives, 
phase locking, jamming, and new types of ratchet effects occur. 
In some regimes we find that the 
addition of a non-ratcheting species
{\it increases} the velocity of the ratcheting particles.
We show that these effects occur due to the collective interactions
of the colloids.  
\end{abstract}
\maketitle

\vskip2pc
 
The motion and ordering of  
colloidal particles on two-dimensional (2D) periodic substrates
has been attracting growing interest
due to recent experimental
breakthroughs that permit the creation of static 1D and 2D
periodic substrate arrays using optical and holographic techniques
\cite{Grier,Bechinger,Reichhardt,Korda,Ladavac,Korrda,Dholakia,Gluckstad}.
It is also possible to create {\it dynamic} periodic
arrays, such as flashing
or shifting traps \cite{Grier,Koss,Ratchet,Babic}.   
Colloidal particles interacting 
with periodic substrates 
are ideal model systems for
studying general problems in condensed matter physics, such 
as the ordering and melting of commensurate and incommensurate 
elastic lattices on periodic surfaces. 
Problems of this type can arise 
in vortex crystal ordering in superconductors \cite{Harada} 
or Bose-Einstein condensates (BECs) with periodic 
pinning sites \cite{Bigelow}
as well as molecules adsorbed on surfaces \cite{Berlinsky}.
The dynamics of colloids moving over periodic substrates 
is also relevant to understanding 
depinning phenomena \cite{Korrda,Olson} 
and phase locking \cite{Korda}. 
In addition to the scientific interest in these systems, 
there are technical applications for colloids moving over periodic
arrays, such as the fractionation or segregation  of
colloidal mixtures 
\cite{Korda,Ladavac,Dholakia,Gluckstad,Koss}. 
For example, species fractionation can be achieved by flowing
colloidal mixtures over an array at an angle  
such that the motion of one 
colloid species
locks to a symmetry direction of the periodic array while
the other species moves 
in the driving direction 
\cite{Korda,Dholakia,Gluckstad}.
Other 
segregation phenomena occur when the 
substrate is dynamic, and
rectification or ratchet devices can be constructed 
in which one species ratchets 
at a different velocity 
than the other 
\cite{Ratchet}. 
New ratchet-based logic devices 
have also been realized \cite{Babic}
which may be useful for understanding similar
solid state nanoscale devices \cite{RCA}.  

Assemblies of interacting particles have been extensively
studied for systems with random disorder and periodic substrates. 
Collectively interacting particles 
on a flashing substrate is 
a new class of problem that 
can be realized experimentally using dynamic 
substrates. 
For particles interacting with optical traps, flashing can be achieved 
simply by modulating the laser power.
It should 
be possible to create similar
flashing potentials for vortices in BECs or ions interacting with optical
arrays \cite{Bigelow}. 
  
In this work we consider mono- and bidisperse assemblies
of charged colloids interacting with flashing 2D symmetric periodic
substrates.  When a dc drive is applied to this system, phase locking
in the velocity force curves occurs.
If an external ac drive is applied instead of a dc drive,
we find that new types of ratchet effects can be realized.
For bidisperse colloidal assemblies, 
the relative velocity between the two 
colloid species is affected by the density, 
number ratio, and charge disparity of the species. 
We find that collective effects play an important role in the ratcheting
behavior.
For a system containing a nonratcheting species $A$, the introduction of
a second species $B$ that does ratchet can induce ratcheting in species
$A$ due to the interactions between the colloid species.
The effectiveness of the induced ratcheting goes through a
peak as the fraction of ratcheting particles is changed at fixed
colloid density.
At high densities, the colloids reach a {\it jammed} state and both
species ratchet at the same velocity. 
We also find a remarkable phenomenon for strongly
interacting particles, where it is possible to {\it increase} the drift
velocity of a ratcheting species by adding non-ratcheting colloids to the
system. 
This effect occurs when the disorder introduced by the non-ratcheting
species breaks apart structures formed by the ratcheting species and
allows the latter to couple more effectively to the substrate. 
Our results can be tested experimentally for
colloids interacting with flashing optical traps.  

We consider a 2D Brownian dynamics (BD) simulation of  
$N$ interacting colloids with periodic boundary conditions 
in both the $x$ and $y$-directions. 
We neglect hydrodynamic and excluded volume interactions, which 
are reasonable assumptions for strongly charge-stabilized particles in the 
low volume fraction limit.
The overdamped equation of motion \cite{Ermack}
for colloid $i$ 
is given by
\begin{equation}
\eta\frac{ d{\bf R}_{i}}{dt} = {\bf F}_{i}^{cc} + {\bf F}^{T}_{i} + 
{\bf F}^{s}_{i} + {\bf F}^{ext}_{i} 
\end{equation}
where the damping constant $\eta$ is set to
unity. The colloid-colloid interaction force is 
${\bf F}_{i}^{cc} = q_{i}\sum^{N_{i}}_{i\neq j}-\nabla_i V(r_{ij})$,
where the colloid-colloid potential is a Yukawa or 
screened Coulomb 
interaction with the form
$V(r_{ij}) = E_0(q_{j}/|{\bf r}_{i} - {\bf r}_{j}|)\exp(-\kappa|{\bf r}_{i} 
- {\bf r}_{j}|)$. Here $E_0=Z^{*2}/(4\pi\epsilon\epsilon_0)$ where 
charge is measured in units of $Z^*$ and $\epsilon$
is the solvent dielectric constant.  Lengths are measured in units
of $a_0$, assumed of order a micron, while time and force are
measured in units of
$\tau=\eta/E_0$
and $F_0=E_0/a_0$. 
$q_{j(i)}$ is the charge on particle $j$($i$), 
$1/\kappa$ is
the screening length which is set to $2a_0$,
and ${\bf r}_{i(j)}$ is the position of particle
$i$($j$). For monodisperse assemblies, $q_{i}/q_{j} = 1$, while
for bidisperse assemblies, $q_{i}/q_{j} \neq 1$.    
The thermal force ${\bf F}^T$ is modeled as
random Langevin kicks 
with the properties $\langle{\bf F}^{T}_{i}\rangle = 0$ and
$\langle{\bf F}^{T}(t){\bf F}^{T}(t^{\prime})\rangle = 2\eta k_{B}T\delta(t - t^{\prime})$. 
The substrate force
${\bf F}^{s}_{i} = A\sin(2\pi x/a_x){\bf {\hat x}} + B\sin(2\pi y/a_y)
{\bf {\hat y}}$ 
with $N_p$ minima
is flashed on and off with a square wave that has a 50\% duty cycle beginning
in the off state at $t=0$.
Here $a_x=4a_0$ is the substrate periodicity, $a_y=\sqrt{3}a_x/2$, 
$A=B$ is the strength of the
substrate, 
$T_s$ is the flashing period, 
and $\omega_s=2\pi/T_s$. 
For most of this work, the sample size is $76 a_0 \times 65.8 a_0$.
The term ${\bf F}^{ext}_{i}$ represents an externally applied
drive. For a dc drive,
${\bf F}^{ext}_{i} = F^{dc}{\bf {\hat x}}$, and for an
ac drive of frequency $\omega_d$, phase offset $\phi$, and no
force off set from zero,
${\bf F}^{ext}_{i}  = F^{ac}\sin((\omega_{d}+\phi)t){\bf {\hat x}}$.
We allow the system to reach a steady state, which takes less than 5000
BD steps, before
measuring the time-averaged particle velocities of each species,
$\langle V_{A(B)}\rangle =(\tau_{av}N_{A(B)})^{-1}\sum_{i=1}^{N_{A(B)}}
\sum_{t=0}^{\tau_{av}}
{\hat {\bf x}}\cdot {\bf v}_i(t)$, 
where $\tau_{av} > T_s$,
$\tau_{av}=$ 5000 to $1 \times 10^6$ BD steps.

\begin{figure}
\includegraphics[width=3.3in]{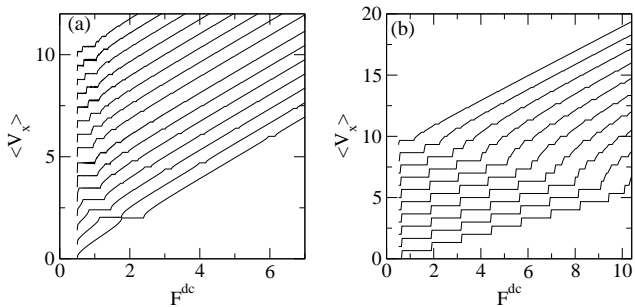}
\vspace{-0.2in}
\caption{
(a) $\langle V_{x}\rangle$ vs $F^{dc}$ for a system with $N/N_{p} = 1$ 
and $N_p=361$ for 
$A=1$ at varied flashing period.
From bottom to top, 
$T_s=100$, 150, 200, 250, 300, 350, 400, 450, 500, 550, 600, 650, 700,
750, and 800 BD steps.
(b) Same as in (a), with a fixed $T_s=300$ and increasing substrate 
strength $A$.  From top to bottom, $A=1$, 2, 3, 4, 5, 6, 7, 8, 9,
and 10.  All curves 
have been vertically offset for
clarity.
\vspace{-0.25in}
}
\end{figure}

We first consider the case of monodisperse colloidal assemblies interacting
with a flashing substrate and a dc drive, where we find
phase locking. 
For a particle moving 
over a periodic substrate with an additional ac drive, phase locking
takes the form of Shapiro steps \cite{Shapiro}, in which 
the particle velocity is periodically 
modulated by the dc motion over the substrate at a frequency 
$\omega_m=a_x\eta/(F^{dc}\tau)$. 
When $\omega_m=n\omega_d$, with $n$ an integer, 
a synchronization of frequencies occurs and the particle 
motion stays locked at the same dc velocity over a range of dc drive,
producing steps in the velocity-force curve. 
In our system we consider a dc drive and a flashing substrate, and there 
is no external ac driving force. The synchronization occurs as 
$\omega_m=n\omega_s$.
Here, the flashing substrate permits the realization of phase locking 
without an external ac drive.

In Fig.~1(a) we show a series of velocity-force curves for varied 
$\omega_s$ for the case $N=N_p$. A depinning threshold appears at 
$F^{dc}=0.34$ followed by a series of 
phase locking steps.
For higher $\omega_s$, steps at higher $F^{dc}$ can be resolved. 
In Fig.~1(b) we plot the velocity force curves for varied substrate strength 
$A$ at fixed $T_s=300$. 
Larger numbers of steps that have a saturated width
appear for larger $A$. 
These saturated steps
are distinct 
from Shapiro steps, where the width of 
the $n$th step varies with ac amplitude as the $n$th-order Bessel function.
For $F^{dc}>A$, particles are no longer trapped in the minima when the
substrate cycles on, and much weaker Shapiro steps appear.
When the number of colloids is incommensurate with 
the substrate, the steps become smooth and have a finite slope as 
additional soliton type pulses of motion are present which do not lock
to $\omega_s$.

\begin{figure}
\includegraphics[width=3.3in]{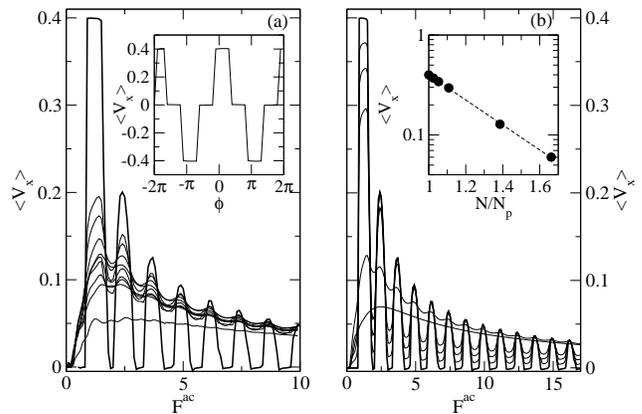}
\vspace{-0.2in}
\caption{
(a) $\langle V_{x}\rangle$ 
vs $F^{ac}$ for a system with $\omega_{s}/\omega_{d} = 1$
and $N_p=64$. 
From top to bottom maximum, $N/N_{p}=1$,
1.25, 1.28, 1.33, 1.375, 1.41, 1.44, 1.47, 1.5, and 2.0.
Inset: $\langle V_{x}\rangle$ for the same system in (a)
for varied phase difference $\phi$ between the ac drive and flashing
substrate. (b) 
Same as in (a) for a higher $N_p=361$ with $N/N_{p}=1$ (top peak), 
1.025, 1.05, 1.10, 1.39, and 1.66
(bottom smooth curve). 
Inset: the maximum value of $\langle V_{x}\rangle$ 
vs $N/N_p$ for the system in (b).  
\vspace{-0.25in}
}
\end{figure}

\begin{figure}
\includegraphics[width=3.4in]{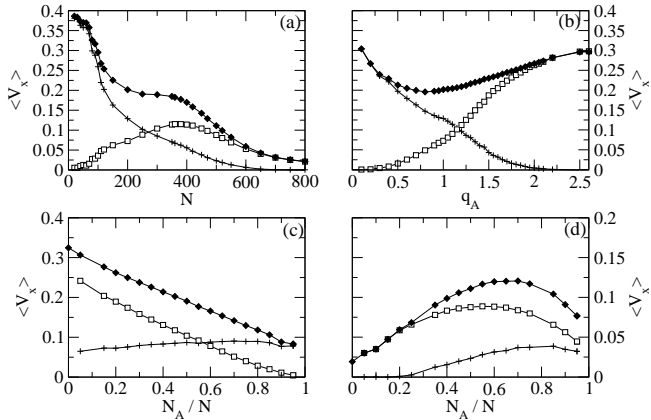}
\vspace{-0.2in}
\caption{
Ratcheting effect for colloidal mixtures in a sample with $N_p=722$. 
Open squares: $\langle V_x^A\rangle$ for
non-ratcheting species $A$;
filled diamonds: $\langle V_x^B\rangle$ for 
ratcheting species $B$; 
plus symbols: relative velocity.
(a) Fixed 
$N_{A}/N = 1/2$ 
for increasing 
$N$. 
(b) Fixed 
$N_{A}/N = 1/2$, 
$N=200$, and $q_B=3$ 
for increasing $q_{A}$.
(c) Fixed  $N = 300$ and increasing $N_{A}/N$. 
(d) Fixed $N = 500$ and increasing $N_{A}/N$.   
\vspace{-0.25in}
}
\end{figure}

If an ac drive is applied instead of a dc drive, it is possible
to produce a ratchet effect.
Consider the case where 
$F^{ac}<A$, such that if the substrate potential were static, 
the particle would remain trapped within a
single pinning plaquette.
The particle is able to move under the influence of the ac drive during
the period of time when the substrate flashes off, but will
be trapped in whichever plaquette it reached when the substrate flashes
back to the on state.  Depending on the amplitude, frequency
$\omega_d$, and phase
of the ac drive relative to the flashing of the substrate, the particle
may be able to move forward a distance $na$ in the $x$ direction with each
cycle of the 
pinning, where $n$ is an integer, 
but is trapped inside a pin during part or all the period
of time when it would otherwise have moved a distance $-na$,
resulting in a net dc motion.
In Fig.~2(a) we show the average 
velocity in the $x$ direction $\langle V_x\rangle$ vs 
$F^{ac}$ for a system with $\omega_{s}/\omega_{d} = 1$, $N_p=64$, 
and $\phi=0$. 
For the commensurate 
case of $N=64$,
the effective ratchet velocity goes 
through a 
series of peaks 
of
decreasing height, with the maximum peak occurring at 
$F^{ac}= 1.25$.  Between the peaks, $\langle V_{x}\rangle \approx 0$.
The decreasing envelope of peak heights 
with increasing $F^{ac}$
arises 
since the particles are able
to depin during part of the cycle when the pins have flashed to the on state,
decreasing the net rectification.
At incommensurate fillings $N\ne N_p$, the colloid positions
are more disordered due to the colloid-colloid interaction forces 
and the rectification peaks become smeared.
In Fig.~2(b) we plot results from a larger system
of $19\times 19$ pins at the same $a_x.$
At $N/N_p=1$, peaks again appear as a function of $F^{ac}$, while 
at the higher colloid densities the peaks are smoothed. 
The peak height is constant 
for $N/N_p\le 1$
since at submatching densities, particle-particle interactions do not
play an important role and the rectification that occurs is a
strictly single-particle process. 
In the inset of Fig.~2(b) we show the monotonic decrease for the maximum peak
value of $\langle V_{x}\rangle$ for increasing $N/N_p$, which
can be fitted well to an exponential decay.
The ratchet effect also depends on the
phase $\phi$ between the flashing pinning and ac driving force. 
In the inset of Fig.~2(a) we    
show $V_{x}$ vs $\phi$ for a fixed $F^{ac} = 1.0$. 
$V_x$ is positive
for $\phi=0$, zero near $\phi=\pi/2$, and negative for $\phi=\pi$. 

\begin{figure}
\includegraphics[width=3.05in]{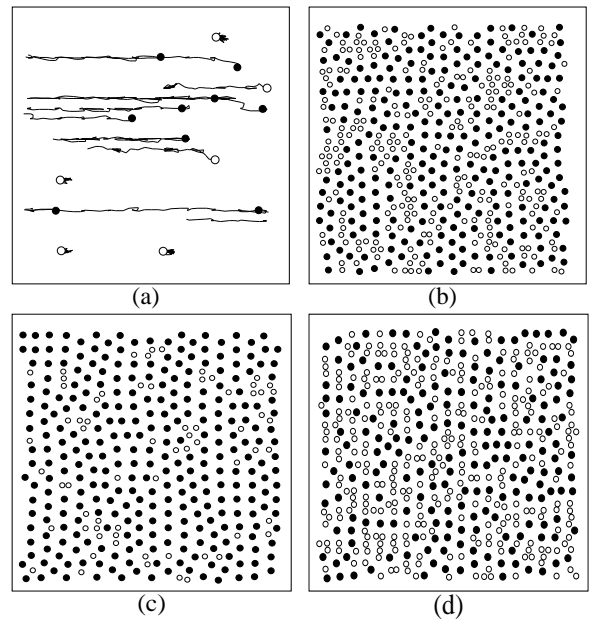}
\vspace{-0.2in}
\caption{
Snapshot of colloid positions (circles) and trajectories (lines) 
in a portion of the sample for 
a fixed period of time for the system in Fig.~3(a) at $N = 40$.
Dark circles are the ratcheting species $B$ and light circles are
the non-ratcheting species $A$. 
(b) Snapshot 
of the jammed
system in Fig.~3(a) at $N = 700$. 
(c) Snapshot at $N_A/N=0.2$ in Fig.~3(d). (d) Snapshot
at $N_A/N=0.6$ in Fig.~3(d).
\vspace{-0.25in}
}
\end{figure}

We next consider mixtures of two colloidal species with different charge.
The coupling to the ac drive is weaker 
for the species with smaller charge.
We first show how the 
density of the system affects the ratcheting
for an ac drive at which species $B$ with charge $q_B=3$ ratchets and 
species $A$ with charge $q_A=1$ does not. 
In Fig.~3(a) we plot 
the average velocities $\langle V_{x}^A\rangle$ and 
$\langle V_{x}^B\rangle$ vs N, 
with fixed 
$N_{A}/N = 1/2$, 
$A = 1.0$, $T_s=T_d=500$,
and $q_{B}/q_{A} = 3$. For low densities, species $B$ 
moves at a finite velocity while $\langle V_x^A\rangle$ is close to zero. 
As $N$ increases, the particles interact
more strongly and the ratcheting particles $B$ push the $A$ particles
forward, 
inducing a ratcheting effect for species $A$.
In Fig.~4(a) we illustrate the motion of the colloidal mixtures at a 
low density. 
All of the species $B$ particles are mobile, as indicated by the
trajectories.
In several places species $A$ particles can be
seen to be immobile. 
However, in the middle and right portions of the figure, two
$A$ particles 
have trails indicating motion that was produced
as a species $B$ particle pushed the species $A$ particle.
This is the mechanism which causes the ratcheting in species $A$. 
We note that if only species $A$ is present, no ratchet effect occurs
for any value of $N$ unless $F^{ac}$ is increased. 
The removal of the last few $B$ particles 
slows the system in a non-linear fashion until the whole system 
stops at $N_A/N=1$. 

As the total number of particles increases for fixed 
$N_A/N=1/2$,
shown in Fig.~3(a), 
$\langle V_x^B\rangle$ drops as
more of the ratcheting energy is transfered to
species $A$. 
We find a maximum in 
$\langle V_x^A\rangle $ at $N\approx 400$. 
The velocity of both species decreases for $N>400$.
For $N > 700$ the system jams and the particles can no longer exchange
positions,  
so both species move with the
same velocity.
In Fig.~4(b) 
at $N = 700$, 
the species 
tend to clump together, with species 
$B$ tending to form crystallites.  
The jamming or freezing transition we find
is similar to the freezing at increased density observed in systems of two 
species of particles moving in different directions \cite{Helbing}. 

In Fig.~3(a), 
the relative velocity $\langle V_x^B\rangle-\langle V_x^A\rangle$ between 
species $A$ and $B$
is a measure of the fractionation, and it drops to
zero when the system jams. In Fig.~3(b)
we plot $\langle V_x^{A}\rangle$ and 
$\langle V_x^{B}\rangle$ vs varied charge $q_{A}$  
for the same system in Fig.~3(a) at fixed density with $N=200$.   
Here $\langle V_x^{A}\rangle$ increases monotonically 
with $q_{A}$, and for $q_{A} > 2.5$, the velocities of
the two species are the same. 
$\langle V_x^B\rangle$ decreases with increasing $q_A$
for $q_{A} < 1.0$ since species $A$ needs to be pushed in order
to ratchet for these values of $q_A$, and the coupling between
the species increases with $q_A$, resulting in a larger drag on
species $B$.  
For $q_{A} > 1$, the $A$ species starts to ratchet on its own
and 
$\langle V_x^B\rangle$ recovers as the drag 
from species $A$ diminishes.   

In Fig.~3(c) for a low density at $N = 300$ 
we show $V_{x}$ as a function of 
$N_A/N$.
As 
the fraction of ratcheting colloids 
goes to zero, 
the velocity of both species is monotonically reduced.
In the low density regime the system is always liquidlike.  
In Fig.~3(d) we show
a very different behavior at
a much higher $N=500$. 
Here, increasing 
the fraction of non-ratcheting particles 
$N_A/N$ {\it increases} the
average velocity of {\it both} the ratcheting and 
non-ratcheting species. 
The relative velocity between the two species also increases
for higher $N_{A}/N$.   
These effects are more pronounced for higher $N$. 
For small
$N_{A}/N$
at high densities,
the colloid-colloid interactions dominate and the system acts 
as a rigid unit so that all the colloids must ratchet at the same 
velocity together, as in Fig.~4(c).
In general some colloids fall at locations that are incommensurate with 
the substrate, so they do not ratchet as effectively and slow 
the entire system. In Fig.~3(d) for low 
$N_{A}/N$, 
$\langle V_x^{A}\rangle = \langle V_x^{B}\rangle$, 
indicating that the system is moving 
elastically.  For higher fractions 
the velocities differ, indicating that tearing of the lattice 
occurs, and at the same time the velocities of both species increase. 
We observe that in this case the system is much more disordered and the
colloids act more as single particles rather than as a collective unit. 
Thus, the addition of the non-ratcheting particles effectively 
breaks up the system or adds disorder, 
as in Fig.~4(d),
which allows the system to behave as a liquid rather
than a solid. 
This improves the coupling of the ratcheting colloids to the substrate since
they are no longer held in incommensurate positions, and increases the
overall effectiveness of the ratcheting.
We have also performed extensive simulations for other parameters
such as varied ratios of $A$ and $B$, 
changing the steepness of the substrate potentials,
densities, $N_{A}/N$, and $T$. 
In all cases we find the same general features of the ratchet
effects described here, leading us to believe that the effects are 
very generic.

To summarize, we have shown that a variety of dynamical behaviors, including 
phase locking, ratchet effects, and jamming, are possible for colloids 
and colloidal mixtures interacting with
flashing symmetric periodic substrates. 
For a flashing substrate with a dc drive,
a phase locking occurs
that is distinct from the typically
studied Shapiro step phase locking. 
If a strictly ac drive is applied
to a flashing substrate, a ratchet effect can occur, with a
ratcheting direction that depends on the relative phase of the two ac 
signals. For mixtures of particles
in regimes where one species ratchets and the other does not, 
we show that 
the ratcheting species can induce a ratchet effect in the non-ratcheting 
species, at the cost of a reduced
effectiveness of ratcheting in the first species.
For high densities the system jams and both species move at the same 
velocity.   
We have demonstrated the ratchet effect
for a wide range of parameters for the mixtures. In the dense regime we
find 
that adding 
nonratcheting colloids can {\it increase}
the drift velocity of the ratcheting species. This is due to the non-ratcheting
particles breaking up the elastic flow of the ratcheting species and 
improving the coupling to the substrate.
Our results should be readily testable 
for colloids on dynamical optical traps.  

This work was supported by the US Department of Energy under Contract No.
W-7405-ENG-36. BJ was supported by the NSF-NIRT Grant No. DMR-02-10519
and by the Alfred P. Sloan Foundation.

\end{document}